\documentclass[twocolumn,showpacs,preprintnumbers, prl,
               superscriptaddress,nofootinbib]{revtex4}
\usepackage{graphicx, fancybox}
\usepackage{amsmath,amssymb,bm}

\begin{document}

\title{
Nonequilibrium time evolution 
of higher order cumulants of conserved charges 
and event-by-event analysis
}

\author{Masakiyo Kitazawa}
\email{kitazawa@phys.sci.osaka-u.ac.jp}
\affiliation{
Department of Physics, Osaka University, Toyonaka, Osaka 560-0043, Japan}

\author{Masayuki Asakawa}
\email{yuki@phys.sci.osaka-u.ac.jp}
\affiliation{
Department of Physics, Osaka University, Toyonaka, Osaka 560-0043, Japan}

\author{Hirosato Ono}
\email{ono@phys.sci.osaka-u.ac.jp}
\affiliation{
Department of Physics, Osaka University, Toyonaka, Osaka 560-0043, Japan}

\begin{abstract}

We investigate the time evolution of higher order cumulants of 
conserved charges in a volume with the diffusion master equation.
Applying the result to the diffusion of non-Gaussian 
fluctuations in the hadronic stage of relativistic heavy ion collisions, 
we show that the fourth-order cumulant of net-electric charge 
at LHC energy is suppressed compared with the recently 
observed second-order cumulant at ALICE, if the 
higher order cumulants at hadronization are suppressed
compared with their values in the hadron phase in equilibrium.
The significance of the experimental information on the rapidity 
window dependence of various cumulants in investigating 
the history of the dynamical evolution of the hot medium
created in relativistic heavy ion collisions is emphasized.

\end{abstract}

\date{\today}

\maketitle

\section{Introduction}

Statistical mechanics tells us that observables 
are fluctuating even in equilibrated medium.
Because the fluctuations are determined by the microscopic 
nature of the medium and sensitive to critical phenomena,
they can be exploited to reveal and characterize 
properties of the medium.
In experimental attempts to map the global nature 
of QCD phase transition at nonzero baryon density 
in relativistic heavy ion collisions, 
fluctuation observables, especially those of 
conserved charges, are believed to be promising
observables to diagnose the property of the hot medium
\cite{Koch:2008ia,Stephanov:1998dy,Asakawa:2000wh,
Jeon:2000wg,Ejiri:2005wq,Stephanov:2008qz,Asakawa:2009aj,
Friman:2011pf}.
Active investigation in heavy ion collisions by 
event-by-event analyses has recently been 
performed at the Relativistic Heavy Ion Collider (RHIC) 
and the Large Hadron Collider (LHC) \cite{STAR,ALICE,PHENIX}.
Numerical analyses of higher order cumulants 
in equilibrium have been also carried out in lattice QCD 
Monte Carlo simulations \cite{lattice}.

As an experiment in which fluctuations are measured, 
heavy ion collisions have several notable features.
First, higher order cumulants, which characterize
the non-Gaussian nature of fluctuations, have 
been experimentally observed with good statistics
up to the fourth order \cite{STAR,PHENIX}.
The measurement is possible because the 
system observed in the experiments is not large; 
the observed particle number is at most of order $10^3$, 
while the event-by-event statistics exceed $10^7$.
Second, the fluctuations observed in experiments, 
especially those of conserved charges, are not 
necessarily the same as those in an equilibrated medium, because 
of the dynamical nature of the hot medium created 
by heavy ion collisions.
Because of these properties, an appropriate 
description of the dynamical evolution of non-Gaussian 
fluctuations is required in order to understand the 
experimental results on higher order cumulants.

Concerning the second point, we remark that 
the recent experimental result on the net-electric charge 
fluctuation, $\langle (\delta N_{\rm Q}^{\rm(net)})^2 \rangle$, 
by ALICE Collaboration at LHC \cite{ALICE} supports 
the non-thermal nature of the observed fluctuation. 
The value of $\langle (\delta N_{\rm Q}^{\rm(net)})^2 \rangle$ 
in this experiment is suppressed compared with the 
one in the equilibrated hadronic medium
which has been calculated by lattice QCD simulations 
\cite{lattice} and the hadron resonance gas (HRG) model 
\cite{Karsch:2010ck}.
Moreover, the dependence of 
$\langle (\delta N_{\rm Q}^{\rm(net)})^2 \rangle$ 
on the size of the rapidity window, $\Delta\eta$, 
shows that the suppression of 
$\langle (\delta N_{\rm Q}^{\rm(net)})^2 \rangle$ 
is more pronounced for larger $\Delta\eta$.
These experimental results are reasonably explained if 
one attributes the suppression to the survival of 
fluctuations generated in the primordial deconfined medium 
\cite{Koch:2008ia,Asakawa:2000wh,Jeon:2000wg,Shuryak:2000pd}.
The charge fluctuation normalized by the total number 
of charged particles, $\langle (\delta N_{\rm Q}^{\rm(net)})^2 \rangle
/ \langle N_{\rm ch} \rangle$, is known to take 
$2\sim3$ times smaller value in the equilibrated deconfined 
medium than the hadronic one 
\cite{Asakawa:2000wh,Jeon:2000wg}.
After hadronization, this small fluctuation 
approaches the equilibrated value in the hadronic medium.
Since the variation of the local density of 
a conserved charge is achieved only through diffusion,
the approach of the fluctuation to the equilibrated value 
becomes slower as the volume where the charge is counted becomes larger.
$\langle (\delta N_{\rm Q}^{\rm(net)})^2 \rangle$ thus 
takes a smaller value as $\Delta\eta$ becomes larger, 
which is consistent with the experimental result at ALICE.

On the other hand, the value of 
$\langle (\delta N_{\rm Q}^{\rm(net)})^2 \rangle$ observed 
at RHIC energy is consistent with the one in the 
equilibrated hadronic medium \cite{RHIC-Q,NMBA}.
The difference between RHIC and LHC energies 
indicates that the evolution of the fluctuation in the 
hot medium is qualitatively different between these energies.

In order to confirm the validity of the above explanation 
on the fluctuation measured at ALICE and clarify the origin 
of the qualitative difference between the experimental results 
at RHIC and LHC energies, it should be instructive to measure 
the $\Delta\eta$ dependences of various other fluctuation observables 
in addition to $\langle (\delta N_{\rm Q}^{\rm(net)})^2 \rangle$ 
in these experiments.
For example, the net-baryon number fluctuation, 
$\langle (\delta N_{\rm B}^{\rm(net)})^2 \rangle$, 
is an experimentally-observable conserved charge 
fluctuation \cite{KA1,KA2}, although neutral baryons are 
not directly observable.
Because the diffusion of the baryon number in the hadronic 
phase is slower than that of the electric charge due to the 
large mass of its carriers, baryons, if the origin of the 
suppression of $\langle (\delta N_{\rm Q}^{\rm(net)})^2 \rangle$ 
at ALICE is indeed traced back to the smallness of
the primordial fluctuations, 
$\langle (\delta N_{\rm B}^{\rm(net)})^2 \rangle$ must have 
steeper suppression as a function of $\Delta\eta$
than $\langle (\delta N_{\rm Q}^{\rm(net)})^2 \rangle$.

In event-by-event analyses, one can also measure 
higher order cumulants of conserved charges 
\cite{STAR,PHENIX} such as the fourth-order ones 
$\langle (N_{\rm Q}^{\rm(net)})^4 \rangle_{\rm c}$ and 
$\langle (N_{\rm B}^{\rm(net)})^4 \rangle_{\rm c}$.
The experimental analysis of these observables as functions 
of $\Delta\eta$ can obviously provide us more information 
on the time evolution of fluctuations in the hot medium.
So far, however, systematic studies on the dynamical evolution of 
higher order cumulants in heavy ion collisions, whose results 
can be compared with their experimental observation, have not 
been carried out to the best of the authors' knowledge.
The purpose of the present Letter is to make the first 
investigation on this issue using a simple but
theoretically lucid model, and to make a prediction 
on the $\Delta\eta$ dependence of higher order 
cumulants in relativistic heavy ion collisions.

\section{Stochastic formalism to describe diffusive systems}

In relativistic heavy ion collisions with sufficiently
large collision energy per nucleon, $\sqrt{s_{\rm NN}}$, 
the hot medium created at mid-rapidity has 
an approximate boost invariance.
Useful coordinates to describe such a system are the 
coordinate-space rapidity $\eta$ and proper time $\tau$.
We denote the net number of a conserved charge
per unit coordinate-space rapidity as $n(\eta,\tau)$.
In a class of experiments, event-by-event fluctuations of 
the charge at kinetic freezeout in a phase 
space corresponding to the rapidity window determined 
by the experiment are observed.
The phase space approximately corresponds to a finite 
coordinate-space rapidity interval \cite{Koch:2008ia}.
Assuming that the kinetic freezeout takes place at a 
certain proper time $\tau_{\rm fo}$, the 
experimentally-observed conserved-charge number at 
mid-rapidity at RHIC and LHC is given by
\begin{align}
Q(\tau) 
= \int_{-\Delta\eta/2}^{\Delta\eta/2} d\eta n(\eta,\tau)
\label{eq:Q}
\end{align}
at $\tau = \tau_{\rm fo}$ with the rapidity window of the detector $\Delta\eta$. 
In the following, we investigate the time evolution of 
higher order cumulants of $Q(\tau )$.

In a sufficiently large space-time scale where 
hydrodynamic equations at first order are applicable, 
the average of $n(\eta,\tau)$ follows the diffusion equation 
\begin{align}
\partial_\tau n(\eta,\tau) 
= D \partial_\eta^2 n(\eta,\tau) ,
\label{eq:diffusion}
\end{align}
where $D$ is the diffusion constant in this coordinate system.
In a boost-invariantly expanding system,
$D$ receives a factor $\tau^{-2}$ compared with the 
diffusion constant in Cartesian coordinate.
In order to describe fluctuations around the solution of 
Eq.~(\ref{eq:diffusion}), one may employ a stochastic 
model, in which the time evolution of the 
deterministic part satisfies Eq.~(\ref{eq:diffusion}).

A choice of such stochastic models is the theory of hydrodynamic 
fluctuations \cite{Landau,Kapusta:2011gt}, in which 
the hydrodynamic equations are promoted to Langevin equations
with stochastic terms representing fast random forces arising 
from microscopic interactions.
In the equation corresponding to the conservation law of 
a charge, Eq.~(\ref{eq:diffusion}), the derivative 
of the stochastic force, $\partial_\eta \xi(\eta,\tau)$, 
is added to the right-hand side of Eq.~(\ref{eq:diffusion}) 
\cite{Shuryak:2000pd}.
The equation is referred to as {\it stochastic diffusion 
equation}.
Up to Gaussian fluctuations, property of $\xi(\eta,\tau)$
is completely determined by the fluctuation-dissipation
relation, which is obtained from the locality of 
$\xi(\eta,\tau)$ and large time behavior of $n(\eta,\tau)$
\cite{Landau}.
It is known that the stochastic equation determined in 
this way well describes Gaussian fluctuations 
in fluids \cite{Landau}.
However, extension of this formalism to treat 
higher order fluctuations is nontrivial.
There is no unique generalization of the 
fluctuation-dissipation relation to higher orders, or 
no a priori justification of such extensions.

Concerning the difficulty in the description of 
non-Gaussian fluctuations, 
it is worthwhile to note a theorem on Markov process,
which states that stochastic forces in a 
Langevin equation for Markov process are of Gaussian 
when the equation describes stochastic variables which 
are continuous and vary continuously 
\footnote{
This condition needs a brief explanation.
For simplicity, we here assume that the stochastic variable
is one dimensional, denoted by $x$ and $y$. 
Let $P(x,t +\Delta t | y, t)$ be 
the conditional probability of $x$ at time $t+\Delta t$ 
given the system was at $y$ at time $t$. 
When the stochastic variable is continuous, it is shown 
that $P(x,t +\Delta t | y, t)$ satisfies the
Lindeberg condition
$\lim_{\Delta t \rightarrow 0}
\frac{1}{\Delta t} \int_{|x-y|>\varepsilon} 
dx ~P(x,t +\Delta t | y, t) =0,$
for arbitrary $\varepsilon > 0$ \cite{Gardiner}.
This condition is used effectively in the
proof of the Gaussianity. 
Note that, when the 
stochastic variables are discrete, the
Lindeberg condition is obviously violated.
}
\cite{Gardiner}.
Since the standard theory of hydrodynamic fluctuations 
describes a Markov process and the hydrodynamic variables 
are continuous, the theorem demands that $\xi(\eta,\tau)$
be of Gaussian; to allow for nonzero higher order 
correlations of $\xi(\eta,\tau)$, one must relax at least 
one of the two conditions, i.e. Markovian and the 
continuity.

Without the higher order correlation of $\xi(\eta,\tau)$,
all higher order cumulants vanish in equilibrium.
Even in such a formalism, the relaxation of non-Gaussianity
starting from a particular initial condition can be described.
In the physics of fluctuations in relativistic heavy ion 
collisions, however, nonzero higher order cumulants 
in equilibrium play a crucial role.
First, the experimental results obtained so far report 
nonzero higher order cumulants near the equilibrated 
values \cite{STAR,PHENIX,ALICE}.
Second, higher order cumulants are expected to 
{\it increase} toward the equilibrated values in the 
hadronic medium \cite{Asakawa:2000wh,Jeon:2000wg}.
To reproduce these features, the stochastic model 
must obviously has nonzero higher order cumulants 
in equilibrium.

\begin{figure}
\begin{center}
\includegraphics[width=.49\textwidth]{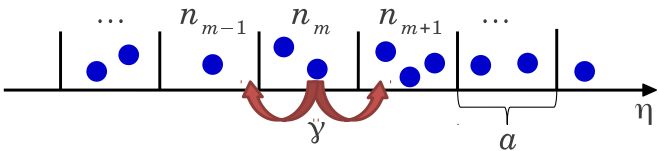}
\caption{
System described by the diffusion master 
equation Eq.~(\ref{eq:DME}).
}
\label{fig:DME}
\end{center}
\end{figure}

In the present study, instead of directly extending 
the theory of hydrodynamic fluctuations, 
we investigate the time evolution of higher order 
cumulants starting from a microscopic model.
In this exploratory analysis, as such a model we consider 
a simple one-dimensional system composed of Brownian particles. 
Instead of tracking the motion of each Brownian particle 
separately, however, we represent the system as follows
(See, Fig.~\ref{fig:DME}).
First, the coordinate $\eta$ is divided 
into discrete cells with an equal length $a$.
Second, we consider a single species of particle for 
the moment, and denote the number of particles in each cell, 
labeled by an integer $m$, as $n_m$, and the probability 
that each cell contains $n_m$ particles as $P(\bm{n},\tau)$ 
with $\bm{n}=(\cdots,n_{m-1},n_{m},n_{m+1},\cdots )$.
Finally, we assume that each particle moves to 
adjacent cells with a probability $\gamma$ per 
unit proper time, as a result of microscopic 
interactions and random motion.
The probability $P(\bm{n},\tau)$ then follows 
the differential equation
\begin{align}
\partial_\tau P(\bm{n},\tau)
=& \gamma \sum_m [
( n_{m} + 1 )  
\{ P(\bm{n}+\bm{e}_{m}-\bm{e}_{m+1},\tau)
\nonumber \\
&+ P(\bm{n} +\bm{e}_{m}-\bm{e}_{m-1},\tau) \} 
\nonumber \\
& -2 n_m P(\bm{n},\tau) ],
\label{eq:DME}
\end{align}
which is referred to as {\it diffusion master equation }
in literature \cite{Gardiner}, 
where $\bm{e}_{m}$ is the vector that all components are
zero except for $m$th one, which takes unity.
We will see later that in the continuum limit, $a\to0$, 
the average density and Gaussian fluctuation of $n(\eta,\tau)$ 
in Eq.~(\ref{eq:DME}) agree with those in 
the stochastic diffusion equation with $D=\gamma a^2$.

Before solving Eq.~(\ref{eq:DME}), some remarks are in order here.
First, Eq.~(\ref{eq:DME}) is often solved with 
an approximation that $n_m$ are sufficiently large 
so that they can be regarded as continuous \cite{Gardiner}.
For the present purpose, however, 
the equation should be solved without this approximation, 
because once $n_m$ are set continuous, 
using an argument similar to the proof of the above-mentioned theorem 
on Markov processes one can show that all higher 
order cumulants become zero in equilibrium \cite{Gardiner}.
Discrete nature of $n_m$ should be kept to give rise to 
non-Gaussianity in equilibrium.
Second, particles described by Eq.~(\ref{eq:DME}) 
behave as Brownian particles \cite{Einstein}
without correlations with one another.
From this property, it is immediately concluded that the 
distribution of $Q$ in the $\tau\to\infty$ limit 
becomes of Poissonian, in which all cumulants take the same 
value $\langle Q^n \rangle_{\rm c}= \langle Q \rangle$,
provided that $\Delta\eta$ is sufficiently narrower than 
the total length of the system.
Note that this behavior of cumulants is consistent
with the hadronic fluctuations in the HRG model, owing to 
the non-interacting nature of hadrons in this model.
Finally, since Eq.~(\ref{eq:DME}) describes a Markov 
process, stochastic effects in this model do not 
have temporal correlations.
This model, therefore, would not be suitable to describe 
non-Gaussian fluctuations enhanced near the critical point, 
where the stochastic forces would have strong 
temporal correlation and thus the process becomes a 
non-Markov one due to critical slowing down.
In the experimental results obtained so far, however, 
significant enhancements of cumulants which could
be attributed to critical phenomena are not observed 
\cite{STAR,PHENIX,ALICE}.
This result implies that critical phenomena do not 
come into play in the diffusion in the hot medium at least 
in the late stage in heavy ion collisions.
In spite of the simplicity of the model Eq.~(\ref{eq:DME}), 
it is thus expected that the model qualitatively 
well describes the time evolution of $\langle Q^n \rangle_{\rm c}$ 
in the hadronic stage.

\section{Solving diffusion master equation}

Now, let us determine the time evolution of cumulants 
for the stochastic process Eq.~(\ref{eq:DME}).
We first consider the time evolution of the probability 
$P(\bm{n},\tau)$ with a fixed initial condition
\begin{align}
P(\bm{n},0) = \prod_m \delta_{n_m,M_m},
\label{eq:fixed}
\end{align}
i.e. the initial particle numbers are fixed 
as $n_m=M_m$ for all $m$
without fluctuation.
By introducing the factorial generating function,
\begin{align}
G_{\rm f} ( \bm{s},\tau )
= \sum_{\bm{n}} \prod_m s_m^{n_m} P( \bm{n},\tau ),
\label{eq:G_f}
\end{align}
one obtains 
\begin{align}
\partial_\tau G_{\rm f} (\bm{s},\tau)
= \gamma \sum_m ( s_{m+1} - 2s_m + s_{m-1} )
\frac{\partial}{\partial s_m} G_{\rm f} (\bm{s},\tau) .
\label{eq:G_f:ME}
\end{align}
Equation~(\ref{eq:G_f:ME}) is a first-order partial
differential equation, and solved with 
the method of characteristics.
The solution with the initial condition 
Eq.~(\ref{eq:fixed}) is given by 
\begin{align}
G_{\rm f} ( \bm{s}, \tau )
= \prod_m \left( \sum_j r_j e^{-\omega_j \tau} e^{-2\pi  jm i/N} \right)^{M_m},
\label{eq:G_f(t)}
\end{align}
with 
\begin{align}
r_j &= \frac1N \sum_{m'} s_{m'} e^{2\pi j{m'} i/N},
\label{eq:FT:ts}
\\
\omega_j &= -\gamma ( e^{2\pi j i/N} + e^{-2\pi j i/N} -2 ),
\end{align}
where $N$ denotes the total number of cells
and $i$ is the imaginary unit.
The factorial cumulants of the Fourier transform of $n_m$, 
$\tilde{n}_k = \sum_m n_m e^{-2\pi  km i/N}$, are given by
\begin{align}
\langle \tilde{n}_{k_1} \tilde{n}_{k_2} \cdots \tilde{n}_{k_l} \rangle_{\rm fc}
=& \left. \frac{\partial^l  K_{\rm f} }
{\partial r_{k_1} \partial r_{k_2} \cdots \partial r_{k_l}}
\right|_{\bm{s}=\bm{1}},
\label{eq:fc}
\end{align}
with $K_{\rm f}(\bm{s},\tau)= \log G_{\rm f} ( \bm{s}, \tau )$.
Using Eqs.~(\ref{eq:fc}) and (\ref{eq:G_f(t)}), 
the factorial cumulants up to the fourth order are calculated to be
\begin{align}
\langle \tilde{n}_k \rangle_{\rm fc}
=& \tilde{M}_k e^{-\omega_k \tau} ,
\label{eq:nk_fc1} \\
\langle \tilde{n}_{k_1} \tilde{n}_{k_2} \rangle_{\rm fc}
=& - \tilde{M}_{k_1+k_2} e^{-(\omega_{k_1}+\omega_{k_2}) \tau} ,
\label{eq:nk_fc2} \\
\langle \tilde{n}_{k_1} \tilde{n}_{k_2} \tilde{n}_{k_3} \rangle_{\rm fc}
=& 2 \tilde{M}_{k_1+k_2+k_3} e^{-(\omega_{k_1}+\omega_{k_2}+\omega_{k_3}) \tau} ,
\label{eq:nk_fc3} \\
\langle \tilde{n}_{k_1} \tilde{n}_{k_2} \tilde{n}_{k_3} 
\tilde{n}_{k_4} \rangle_{\rm fc}
=& -6 \tilde{M}_{k_1+k_2+k_3+k_4} 
\nonumber \\
&e^{-(\omega_{k_1}+\omega_{k_2}+\omega_{k_3}+\omega_{k_4}) \tau} ,
\label{eq:nk_fc4}
\end{align}
with $\tilde{M}_k = \sum_m M_m e^{-2\pi km i/N}$.

The cumulants of $n_m$ are given by 
\begin{align}
\langle n_{m_1} n_{m_2} \cdots n_{m_l} \rangle_{\rm c}
=& \left. \frac{\partial^l  K }
{ \partial \theta_1 \cdots \partial \theta_l } \right|_{\bm{\theta}=0} .
\label{eq:nm}
\end{align}
with $K(\bm{\theta},\tau) = K_{\rm f}(\bm{s},\tau)|_{s_m=e^{\theta_m}}$.
Using Eq.~(\ref{eq:nm}), they are related to 
Eqs.~(\ref{eq:nk_fc1}) - (\ref{eq:nk_fc4}) as 
\begin{align}
\langle \tilde{n}_k \rangle_{\rm c}
=& \langle \tilde{n}_k \rangle_{\rm fc} ,
\\
\langle \tilde{n}_{k_1} \tilde{n}_{k_2} \rangle_{\rm c}
=& \langle \tilde{n}_{k_1} \tilde{n}_{k_2} \rangle_{\rm fc}
+ \langle \tilde{n}_{k_1+k_2} \rangle_{\rm fc} ,
\end{align}
and so forth.

Since we are interested in the solution of Eq.~(\ref{eq:DME}) 
in the continuum limit, $a\to0$, here we introduce the 
shorthand notation for this limit:
the particle density per unit rapidity $n(\eta) = n_m/a$,
with the rapidity of the $m$th cell $\eta=ma$, 
$\omega_p=\gamma a^2p^2$ with the conjugate 
momentum $p = 2\pi k / Na$.
Also, the probability $P(\bm{n},\tau)$ is promoted to 
a functional, which we denote as $P[n(\eta),\tau]$.
Note, however, that these notations are conceptual; 
In actual applications, the functional $P[n(\eta),\tau]$ 
is understood as the limit of the function $P(\bm{n},\tau)$ 
with small but finite $a$.
One finds from Eq.~(\ref{eq:nk_fc1}) that the continuum 
limit has to be taken with fixed $D=\gamma a^2$, so that 
the deterministic part of Eq.~(\ref{eq:DME}) 
follows Eq.~(\ref{eq:diffusion}).
The factorial cumulants of $\tilde{n}(p)$ in the continuum limit
are obtained by simply replacing $k$ with $p$ in 
Eqs.~(\ref{eq:nk_fc1}) - (\ref{eq:nk_fc4}).

In the following, we consider an infinitely long system
without boundaries.
With the fixed initial condition $n(\eta,0)=M(\eta)$, 
the cumulants of Eq.~(\ref{eq:Q}) at proper time $\tau$ 
are calculated to be
\begin{align}
\langle (Q(\tau))^n \rangle_{\rm c} 
= \int_{-\infty}^{\infty} d\eta M(\eta) H^{(n)}_X(\eta) ,
\label{eq:<Q^n>}
\end{align}
with
\begin{align}
H^{(1)}_X(z) =& I_X(z/\Delta\eta) ,
\\
H^{(2)}_X(z) =& I_X(z/\Delta\eta) - I_X(z/\Delta\eta)^2 ,
\\
H^{(3)}_X(z) =& I_X(z/\Delta\eta) - 3 I_X(z/\Delta\eta)^2 + 2 I_X(z/\Delta\eta)^3 ,
\\
H^{(4)}_X(z) =& 
I_X(z/\Delta\eta) - 7 I_X(z/\Delta\eta)^2 + 12 I_X(z/\Delta\eta)^3 
\nonumber \\
&- 6 I_X(z/\Delta\eta)^4 ,
\end{align}
and
\begin{align}
I_X(z) =& \int_{-1/2}^{1/2} dx \int \frac{dq}{2\pi}
e^{-X^2 q^2} e^{iq(x+z)} ,
\label{eq:I_X}
\end{align}
where $\Delta\eta$ and $\tau$ dependences are 
encoded in the dimensionless parameter 
\begin{align}
X= \frac{\sqrt{D\tau}}{\Delta\eta}.
\label{eq:X}
\end{align}

Next, we extend this result to general initial
conditions containing fluctuations.
We also extend the result to the 
system with two particle species with densities 
$n_1(\eta,\tau)$ and $n_2(\eta,\tau)$, and consider 
cumulants of the difference 
\begin{align}
Q_{\rm(net)}(\tau) = \int_{-\Delta\eta/2}^{\Delta\eta/2}d\eta 
(n_1(\eta,\tau)-n_2(\eta,\tau)),
\label{eq:Q_net}
\end{align}
in order to compare the results with the cumulants of 
the net-electric charge and baryon numbers,
which are given by the difference of particle numbers.
When the two particle species separately follow 
Eq.~(\ref{eq:DME}), the probability with the initial 
condition $P[n_1(\eta),n_2(\eta),0]=F[M_1(\eta),M_2(\eta)]$ 
is given by the superposition of the solutions of 
fixed initial conditions 
\begin{align}
\lefteqn{P[n_1(\eta),n_2(\eta),\tau] =}
\nonumber \\
&\sum_{\{M_1(\eta),M_2(\eta)\}} F[M_1(\eta),M_2(\eta)] 
P_{M_1} [ n_1(\eta),\tau ] P_{M_2} [ n_2(\eta),\tau ],
\label{eq:P[n(x)]}
\end{align}
where $P_{M} [ n(\eta),\tau ]$ is the solution of 
Eq.~(\ref{eq:DME}) with the fixed initial condition
$n(\eta,0)=M(\eta)$, and the sum runs over the function 
spaces of $M_1(\eta)$ and $M_2(\eta)$.
From Eq.~(\ref{eq:P[n(x)]}), one finds that the 
cumulants of Eq.~(\ref{eq:Q_net}) 
are obtained by using the cumulants of 
$P_{M} [ n(\eta),\tau ]$ and $F[M_1(\eta),M_2(\eta)]$
using the superposition formula given in Refs.~\cite{KA2ap,prep}.
The general results will be presented in Ref.~\cite{prep}.

In the present Letter, we focus on the results 
for initial conditions satisfying spatial uniformity and 
locality, i.e.
\begin{align}
\lefteqn{
\langle M_{i_1}(\eta_1)M_{i_2}(\eta_2)\cdots M_{i_l}(\eta_l) \rangle_{\rm c}
}
\nonumber \\
&= [ M_{i_1} M_{i_2} \cdots M_{i_l} ]_{\rm c}
\delta(\eta_1-\eta_2) \cdots \delta(\eta_1-\eta_l) ,
\label{eq:locality}
\end{align}
where this equation defines 
$[ M_{i_1} M_{i_2} \cdots M_{i_l} ]_{\rm c}$ 
on the right-hand side, which are generalized 
susceptibilities to higher orders and off-diagonal components.
The condition Eq.~(\ref{eq:locality}) is 
satisfied, for example, in free gas, as well as systems 
well described by hydrodynamic equations, in equilibrium.
With this initial condition,
cumulants of $Q_{\rm(net)}(\tau)$ are given by
\begin{align}
\langle Q_{\rm(net)} \rangle_{\rm c} 
=& \Delta\eta [ M_{\rm(net)} ]_{\rm c} ,
\label{eq:<barQ^1>L}
\\
\langle Q_{\rm(net)}^2 \rangle_{\rm c} 
=& 
\Delta\eta \{ [ M_{\rm(net)}^2 ]_{\rm c} F^{(2)}_X
+ [ M_{\rm(tot)} ]_{\rm c} ( 1 - F^{(2)}_X ) \},
\label{eq:<barQ^2>L}
\\
\langle Q_{\rm(net)}^3 \rangle_{\rm c} 
=& \Delta\eta \{ [ M_{\rm(net)}^3 ]_{\rm c} F^{(3)}_X
\nonumber \\
& + 3 [ M_{\rm(net)} M_{\rm(tot)} ]_{\rm c} ( F^{(2)}_X - F^{(3)}_X )
\nonumber \\
& + [ M_{\rm(net)} ]_{\rm c} ( 1 - 3 F^{(2)}_X + 2 F^{(3)}_X ) \},
\label{eq:<barQ^3>L}
\\
\langle Q_{\rm(net)}^4 \rangle_{\rm c} 
=& 
\Delta\eta \{ [ M_{\rm(net)}^4 ]_{\rm c} F^{(4)}_X
\nonumber \\
& + 6 [ M_{\rm(net)}^2 M_{\rm(tot)} ]_{\rm c} ( F^{(3)}_X - F^{(4)}_X )
\nonumber \\
& + 3 [ M_{\rm(tot)}^2 ]_{\rm c} ( F^{(2)}_X - 2 F^{(3)}_X + F^{(4)}_X )
\nonumber \\
& + 4 [ M_{\rm(net)}^2 ]_{\rm c} ( F^{(2)}_X - 3 F^{(3)}_X + 2 F^{(4)}_X )
\nonumber \\
& + [ M_{\rm(tot)} ]_{\rm c} (1 - 7 F^{(2)}_X + 12 F^{(3)}_X - 6 F^{(4)}_X ) \},
\label{eq:<barQ^4>L}
\end{align}
with 
\begin{align}
F^{(n)}_X = \int_{-\infty}^{\infty} dz [I_X(z)]^n,
\label{eq:F^n}
\end{align}
and $M_{\rm(net),(tot)}(\eta) = M_1(\eta)\mp M_2(\eta)$, respectively.
$F^{(n)}_X$ with $n\ge2$ are monotonically decreasing functions
of $X$ and satisfies $F^{(n)}_0=1$ and $\lim_{X\to\infty} F^{(n)}_X=0$.
From this property of $F^{(n)}_X$ and Eqs.~(\ref{eq:<barQ^1>L}) 
- (\ref{eq:<barQ^4>L}), one easily finds that the 
distribution of $Q_{\rm(net)}$ approaches a Skellam one with 
\begin{align}
\lim_{\tau\to\infty}\langle Q_{\rm(net)}^{2n+1} \rangle_{\rm c} 
&= \Delta\eta [ M_{\rm(net)} ]_{\rm c} ,
\label{eq:QMnet}
\\
\lim_{\tau\to\infty}\langle Q_{\rm(net)}^{2n} \rangle_{\rm c} 
&= \Delta\eta [ M_{\rm(tot)} ]_{\rm c} .
\label{eq:QMtot}
\end{align}
From Eq.~(\ref{eq:<barQ^2>L}), one also finds that 
the time evolution of the Gaussian fluctuation with the initial
condition Eq.~(\ref{eq:locality}) is equivalent with the one 
in the stochastic diffusion equation \cite{Shuryak:2000pd}
with the fluctuation in equilibrium Eq.~(\ref{eq:QMtot}).

The details of the procedure to deal with Eq.~(\ref{eq:DME})
omitted in this Letter will be elucidated in the 
forthcoming publication \cite{prep}.

\section{Diffusion in hadronic medium}

Next, let us investigate the diffusion of higher order 
cumulants in the hadronic medium in relativistic heavy 
ion collisions using the above results.
To make the argument simple, we assume that a boost 
invariant system with local equilibration is realized 
just above the critical temperature of the deconfinement 
transition.
Then, the fluctuations at this time satisfy 
the locality condition Eq.~(\ref{eq:locality}).
The hot medium then undergoes hadronization and chemical 
freezeout, which take place at almost the same time, 
$\tau=\tau_0$, for large $\sqrt{s_{\rm NN}}$.
Because of the local charge conservations, fluctuations of 
local densities of the conserved charges are 
unchanged during these processes \cite{Koch:2008ia}.
The fluctuations of conserved charges just after 
hadronization thus also satisfy Eq.~(\ref{eq:locality})
to a good approximation,
with cumulants in the equilibrated deconfined medium, 
which are significantly smaller than the ones in the
equilibrated hadronic medium 
\cite{Asakawa:2000wh,Jeon:2000wg,Ejiri:2005wq}.
We take this configuration at $\tau=\tau_0$ 
as the initial condition.

Due to the diffusion in the hadronic phase, 
the fluctuations approach the equilibrated values
in the hadronic medium until kinetic freezeout 
at $\tau=\tau_{\rm fo}$.
Provided that the $\tau$ dependence of the diffusion 
constant $D$ in Eq.~(\ref{eq:diffusion}) is weak, 
this diffusion process is approximately 
described by Eq.~(\ref{eq:DME}).

Now we consider the diffusion of the net-electric charge 
and baryon numbers, $N_{\rm Q}^{\rm(net)}$ and 
$N_{\rm B}^{\rm(net)}$, as conserved charges.
Since experimental measurements of odd order cumulants of 
these charges are difficult at LHC energy because of their
smallness, we also limit 
our attention to the second- and fourth-order cumulants.
Because these charges in the hadron phase are predominantly 
carried by charged pions and nucleons, respectively, 
the cumulants of $N_{\rm Q}^{\rm(net)}$ and $N_{\rm B}^{\rm(net)}$ 
at kinetic freezeout are approximately given by 
Eqs.~(\ref{eq:<barQ^2>L}) and (\ref{eq:<barQ^4>L}) 
with $\tau=\tau_{\rm fo}-\tau_0$, where $n_1(\eta,\tau)$
and $n_2(\eta,\tau)$ are the densities of positive and negative pions, and
nucleons and anti-nucleons, respectively.

To see the behavior of the cumulants determined by
Eqs.~(\ref{eq:<barQ^2>L}) and (\ref{eq:<barQ^4>L}),
one must fix the parameters for the initial condition in 
these equations.
As discussed above, 
the cumulants of the conserved charges at $\tau=\tau_0$ 
are suppressed compared with the equilibrated values.
For even $n$, this condition is represented as 
$[ M_{\rm(net)}^n ]_{\rm c} \ll [ M_{\rm(tot)} ]_{\rm c}$, 
because the cumulants in equilibrium are given by 
Eq.~(\ref{eq:QMtot}).
Because of the suppression of 
$[ M_{\rm(net)}^2 ]_{\rm c} $,
unless $(\delta M_{\rm(net)})^2$ and $M_{\rm(tot)}$ 
have a strong positive correlation
$[ M_{\rm(net)}^2 M_{\rm(tot)} ]_{\rm c} \ll [ M_{\rm(tot)} ]_{\rm c}$ 
should also be satisfied.
On the other hand, the value of $[ M_{\rm(tot)}^2 ]_{\rm c}$ 
in Eq.~(\ref{eq:<barQ^4>L}) is not constrained by 
the conservation laws and strongly depends on the
hadronization mechanism. In the following, we treat this
quantity as a parameter that characterizes the hadronization
mechanism, and propose to utilize this parameter to 
constrain the ambiguity in it.

\begin{figure}
\begin{center}
\includegraphics[width=.48\textwidth]{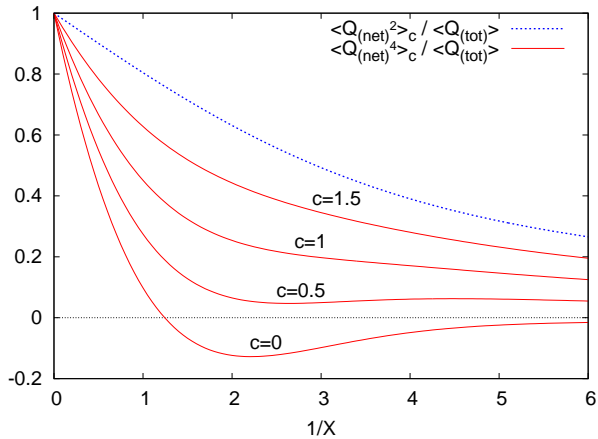}
\caption{
Second and fourth order cumulant of $Q_{\rm(net)}$ as a function
of $1/X$ with the initial condition 
$[ M_{\rm(net)}^2 ]_{\rm c} = [ M_{\rm(net)}^4 ]_{\rm c}
= [ M_{\rm(net)}^2 M_{\rm(tot)} ]_{\rm c} = 0$.
}
\label{fig:Q}
\end{center}
\end{figure}

In Fig.~\ref{fig:Q}, we show the $1/X$ dependence 
of $\langle Q_{\rm(net)}^2 \rangle_{\rm c}$ and 
$\langle Q_{\rm(net)}^4 \rangle_{\rm c}$ given by 
Eqs.~(\ref{eq:<barQ^2>L}) and (\ref{eq:<barQ^4>L}) 
normalized by the equilibrated value Eq.~(\ref{eq:QMtot})
with the initial condition that the fluctuations of 
conserved charges do not exist at all, 
\begin{align}
[ M_{\rm(net)}^2 ]_{\rm c} = [ M_{\rm(net)}^4 ]_{\rm c}
= [ M_{\rm(net)}^2 M_{\rm(tot)} ]_{\rm c} = 0.
\label{eq:<M>=0}
\end{align}
Since the value of $[ M_{\rm(tot)}^2 ]_{\rm c}$ is not
constrained by the local charge conservation at hadronization 
as we discussed above, we regard the ratio 
\begin{align}
c = \frac{ [ M_{\rm(tot)}^2 ]_{\rm c} }{ [ M_{\rm(tot)} ]_{\rm c} },
\label{eq:c}
\end{align}
as a free parameter and 
vary $c$ in the range $0 \le c \le 1.5$.
The result on the second order, 
$\langle Q_{\rm(net)}^2 \rangle_{\rm c}$,
reproduces the one obtained by the stochastic diffusion equation
\cite{Shuryak:2000pd}.

In the figure, one finds that the behaviors of 
$\langle Q_{\rm(net)}^4 \rangle_{\rm c}$ 
and $\langle Q_{\rm(net)}^2 \rangle_{\rm c}$ as functions of 
$1/X$ are qualitatively different.
While the behavior of $\langle Q_{\rm(net)}^4 \rangle_{\rm c}$ 
depends on the value of $c$, 
$\langle Q_{\rm(net)}^4 \rangle_{\rm c}$ is suppressed 
compared with $\langle Q_{\rm(net)}^2 \rangle_{\rm c}$ 
in the parameter range $c\lesssim1.5$.
Note that $\langle Q_{\rm(net)}^4 \rangle_{\rm c}$ can 
become negative in a range of $1/X$, 
although the fourth-order cumulants at the initial time
and in equilibrium are both non-negative.
It is interesting that the sign of the fourth-order
cumulant can be flipped owing to the non-equilibrium effect.

Since $1/X$ is proportional to $\Delta\eta$, 
the result in Fig.~\ref{fig:Q} can directly be compared 
with the experimental result on the $\Delta\eta$ dependence of 
the net-electric charge cumulants 
$\langle (N_{\rm Q}^{\rm(net)})^2 \rangle_{\rm c}
=\langle (\delta N_{\rm Q}^{\rm(net)})^2 \rangle$ and 
$\langle (N_{\rm Q}^{\rm(net)})^4 \rangle_{\rm c}$.
In particular, Fig.~\ref{fig:Q} indicates that 
the $\langle (N_{\rm Q}^{\rm(net)})^4 \rangle_{\rm c}$ at ALICE,
which has not been measured yet, will
be suppressed compared with the 
$\langle (N_{\rm Q}^{\rm(net)})^2 \rangle_{\rm c}$ 
which has been already measured \cite{ALICE}.
This statement is, however, altered for $c\gtrsim2$.
The same conclusion is also anticipated for the relation 
between the baryon number cumulants, 
$\langle (N_{\rm B}^{\rm(net)})^2 \rangle_{\rm c}$ and 
$\langle (N_{\rm B}^{\rm(net)})^4 \rangle_{\rm c}$.

The suppression of the fourth-order cumulant compared with
the second-order one can be intuitively understood as follows.
In our analysis, we consider time evolution with the 
initial condition with small fluctuation. 
With this initial condition, the probability distribution 
becomes wider as $\tau$ becomes larger, and thus 
longer time is required until the probability 
in the tail is equilibrated.
Because the higher order cumulants are sensitive to the 
tail behavior of the probability distribution, 
the approach of the cumulants to the equilibrated values 
is slower for higher order ones.

\begin{figure}
\begin{center}
\includegraphics[width=.48\textwidth]{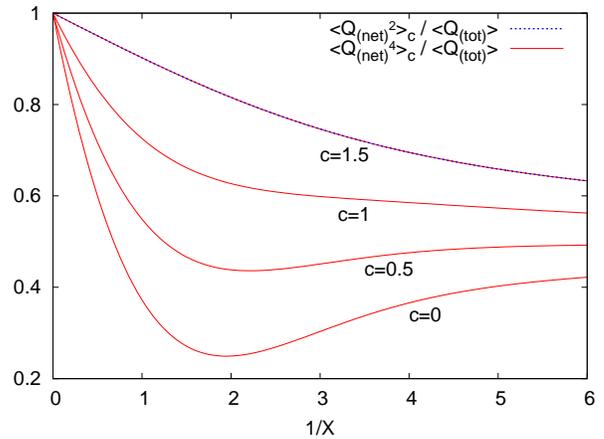}
\caption{
Second and fourth order cumulant of $Q_{\rm(net)}$ as a function
of $1/X$ with $[ M_{\rm(net)}^2 ]_{\rm c}
= [ M_{\rm(net)}^4 ]_{\rm c}
= [ M_{\rm(net)}^2 M_{\rm(tot)} ]_{\rm c}
= 0.5 [ M_{\rm(tot)} ]_{\rm c}$.
}
\label{fig:Q2}
\end{center}
\end{figure}

The result in Fig.~\ref{fig:Q} is obtained with the idealized 
initial condition Eq.~(\ref{eq:<M>=0}), where fluctuations
of the conserved charges completely vanish at $\tau=\tau_0$.
Next, we see how this result is modified when 
these cumulants have small but nonzero
values at $\tau=\tau_0$.
In Fig.~\ref{fig:Q2} we show $1/X$ dependences of 
$\langle Q_{\rm(net)}^2 \rangle_{\rm c}$ and 
$\langle Q_{\rm(net)}^4 \rangle_{\rm c}$ with
the initial condition,  
\begin{align}
[ M_{\rm(net)}^2 ]_{\rm c}
= [ M_{\rm(net)}^4 ]_{\rm c}
= [ M_{\rm(net)}^2 M_{\rm(tot)} ]_{\rm c}
= \frac12 [ M_{\rm(tot)} ]_{\rm c}.
\end{align}
The figure shows that the suppression of 
$\langle Q_{\rm(net)}^4 \rangle_{\rm c}$ compared with 
$\langle Q_{\rm(net)}^2 \rangle_{\rm c}$ is not 
as clear as the result in Fig.~\ref{fig:Q},
but $\langle Q_{\rm(net)}^4 \rangle_{\rm c} < 
\langle Q_{\rm(net)}^2 \rangle_{\rm c}$ is still satisfied 
for a rather wide range of $c$.
The suppression of $\langle Q_{\rm(net)}^4 \rangle_{\rm c}$
thus is a rather robust feature reflecting 
the small initial fluctuations.
Therefore, this behavior of the cumulants at LHC 
energy can be used as an experimental probe to confirm
the suppression of the fluctuations at hadronization.

The results in Figs.~\ref{fig:Q} and \ref{fig:Q2} also 
show that the cumulants $\langle Q_{\rm(net)}^2 \rangle_{\rm c}$ 
and $\langle Q_{\rm(net)}^4 \rangle_{\rm c}$ have characteristic
behaviors as functions of $\Delta\eta$ depending on the 
initial condition.
These results indicate that experimental measurements of
not only the magnitudes of various cumulants at a fixed $\Delta\eta$
but also their $\Delta\eta$ dependence
enable us to explore various information 
on the time evolution of the hot medium and 
the hadronization mechanism in the experiments.
In particular, these analyses would enable us to 
estimate the magnitude of the parameter $c$.
Because this parameter is sensitive to the hadronization 
mechanism, such experimental information would be used 
as an important clue to it.

\section{Discussions}

In this study, we employed a simple model, Eq.~(\ref{eq:DME}), 
to investigate the time evolution of non-Gaussian fluctuations.
The model is expected to describe well the qualitative feature of 
the diffusion of higher order cumulants in the hadronic stage 
as already discussed. In order to explore the diffusion 
in each stage in heavy ion collisions more quantitatively, 
however, one must take various effects into account.
In the present model, for example, fluctuations 
in equilibrium are given by the Poisson or Skellam 
distribution as a result of the absence of interactions 
between each Brownian particle. 
The model thus is not suitable to describe systems 
where equilibrated fluctuations do not follow 
either of these distributions.
The easiest way to treat non-Poissonian fluctuations is 
to consider a system composed of several particle species 
having different charges which separately follow Eq.~(\ref{eq:DME}).
The fluctuations of total charge of all particle species then 
becomes neither Poissonian nor Skellam distribution.
This modification would be used as a first approximation 
to model the non-Skellam behavior of the net-electric 
charge fluctuations.
When the non-Poissonian behavior comes 
from interaction between particles, terms describing 
the interaction should be introduced in Eq.~(\ref{eq:DME}).
Next, the model Eq.~(\ref{eq:DME}) describes a Markov 
process and the stochastic effect does not have 
temporal correlations. 
To take account of nonzero temporal correlations,
the model should be extended to non-Markov ones.
Such a treatment would be required in, for example, dealing with
the non-Gaussianity associated with critical phenomena and 
considering phenomena of the time scale of microscopic interaction.
In the present study, it is also assumed that the numbers of 
two particle species $n_1(\eta,\tau)$ and $n_2(\eta,\tau)$ 
which carry opposite charges are separately conserved.
While this assumption would be well justified for baryon
numbers after chemical freezeout, 
effects of pair-creation and annihilation of electric 
charges would modify the time evolution of 
the net-electric charge fluctuations qualitatively.
Finally, whereas the system investigated in the present study 
corresponds to the one with fixed diffusion constant $D$,
$\tau$ dependence of this parameter should also be taken
into account especially when one investigates the expanding 
system \cite{Kapusta:2011gt}.

Besides these technical issues on the model, several 
assumptions also have been introduced 
on the geometry of the system and initial condition.
To simplify the arguments, in the present Letter 
we limited our attention to the system which is infinitely 
long and the event-by-event configuration of the initial 
distribution is uniform and local, 
and thus satisfy Eq.~(\ref{eq:locality}).
The hot medium created by heavy ion collisions, on the 
other hand, is a finite system, and the effects of 
global charge conservation and the violation of boost 
invariance modify the values of the fluctuation observables
\cite{Koch:2008ia}
and render them dependent on 
the position of the rapidity window.
The validity of locality condition in Eq.~(\ref{eq:locality})
should also be investigated carefully.
To investigate these effects,
the model Eq.~(\ref{eq:DME}) has to be solved with
appropriate boundary conditions and various initial conditions.
Part of these issues will be investigated 
in the forthcoming paper \cite{prep}.

In the present study, we limited our attention to 
sufficiently large $\sqrt{s_{\rm NN}}$, where the 
measurement of odd order cumulants are difficult
owing to their smallness.
On the other hand, in the range of $\sqrt{s_{\rm NN}}$ 
where the beam energy scan program at RHIC is exploring, 
the third order cumulants are also observable 
\cite{STAR,PHENIX} 
and will carry significant information on
the QCD phase transition \cite{Asakawa:2009aj}.
The time evolution and $\Delta\eta$ dependence of 
the third-order cumulants will also be discussed 
in Ref.~\cite{prep}.

In the present Letter, we investigated the time evolution 
of non-Gaussian fluctuations of conserved charges 
in a volume through diffusion using the diffusion master 
equation Eq.~(\ref{eq:DME}), which is intended to model
the time evolution of higher order cumulants of conserved
charges $N_{\rm Q}^{\rm(net)}$ and $N_{\rm B}^{\rm(net)}$ 
in the hadronic stage of relativistic heavy ion collisions.
Our analysis shows that the fourth-order cumulant 
$\langle (N_{\rm Q}^{\rm(net)})^4 \rangle_{\rm c}$
is suppressed compared with the second-order one 
$(\langle N_{\rm Q}^{\rm(net)})^2 \rangle_{\rm c}$, 
provided that the suppression of 
$\langle (N_{\rm Q}^{\rm(net)})^2 \rangle_{\rm c}$
at ALICE \cite{ALICE} is a consequence of 
the survival of the smallness of the fluctuation at the primordial stage.
The same conclusion is also anticipated for baryon 
number cumulants.
It should, however, be noted that this result 
qualitatively depends on the parameter $c$ in 
Eq.~(\ref{eq:c}), 
which is not constrained only by the local charge 
conservation and sensitive to hadronization mechanism.
We emphasize that the fourth-order cumulants of net-electric 
charge and baryon numbers,
$\langle (N_{\rm Q}^{\rm(net)})^4 \rangle_{\rm c}$ and 
$\langle (N_{\rm B}^{\rm(net)})^4 \rangle_{\rm c}$, 
can be observable as functions of $\Delta\eta$ in the 
mid-rapidity region in heavy ion collisions at LHC and RHIC.
The comparison of the experimental data on various cumulants
with the present analysis will reveal the dynamical 
evolution of fluctuations and hadronization mechanism
in heavy ion collisions.

M.~K. acknowledges fruitful discussions at EMMI Rapid Reaction 
Task Force ``Probing the Phase Structure of Strongly 
Interacting Matter with Fluctuations: Theory and Experiment'' 
held at GSI, 
February 11-22, 2013.
He also thanks T.~Kunihiro for discussions.
This work is supported in part by Grants-in-Aid for 
Scientific Research by Monbu-Kagakusyo of Japan 
(Nos.~23540307 and 25800148).

\end{document}